\documentclass[12pt]{article}
\begin{document}
\baselineskip=18.6pt plus 0.2pt minus 0.1pt
\def\be{\begin{equation}}
  \def\ee{\end{equation}}
  \def\bea{\begin{eqnarray}}
  \def\eea{\end{eqnarray}}
  \def\nn{\nonumber\\ }
  \def\RR{\mathbb{R}}
\newcommand{\nc}{\newcommand}
\nc{\al}{\alpha} \nc{\bib}{\bibitem} \nc{\cp}{\C{\bf P}}
\nc{\la}{\lambda}
\nc{\C}{\mbox{\hspace{1.24mm}\rule{0.2mm}{2.5mm}\hspace{-2.7mm}
C}}
\nc{\R}{\mbox{\hspace{.04mm}\rule{0.2mm}{2.8mm}\hspace{-1.5mm} R}}
\begin{titlepage}
\title{
\begin{flushright}
{\normalsize \small
IFT-UAM/CSIC-04-39 }
\\[1cm]
\mbox{}
\end{flushright}
{\bf   On Non Commutative  G2  structure
}\\[.3cm]
\author{A. Belhaj$^{a,b}$\thanks{{\tt adil.belhaj@uam.es}}  and  M. P. Garc\'\i a del
Moral$^{a}$\thanks{{\tt gdelmoral@delta.ft.uam.es}}\\[.3cm] {\it
\small $^{a}$ Instituto de F\'{\i}sica Te\'orica, C-XVI,
Universidad Aut\'onoma de Madrid } \\ {\it \small Cantoblanco,
E-28049-Madrid, Spain}\\
{\it
\small $^{b}$ Departamento  de F\'{\i}sica Te\'orica, C-XI,
Universidad Aut\'onoma de Madrid } \\ {\it \small Cantoblanco,
E-28049-Madrid, Spain}
\\[.2cm]
  } } \maketitle
\thispagestyle{empty}
\begin{abstract}
Using  an  algebraic orbifold  method, we present non-commutative  aspects
of   $G_2$ structure of seven dimensional real manifolds.   We first develop and solve the non
commutativity  parameter constraint
equations
defining      $G_2$  manifold   algebras.
We  show  that there are eight possible solutions  for this extended structure,
one of which corresponds to the commutative case.  Then we obtain
a matrix representation solving such algebras using combinatorial
arguments. An application to matrix model of M-theory is
discussed.
\end{abstract}
\end{titlepage}
\newpage
\section{Introduction}
It has been known for a long time that non-commutative (NC)
geometry plays an interesting role in the context of string theory
\cite{W} and, more recently, in certain compactifications of the
Matrix formulation of M-theory on NC tori
\cite{CDS}. These studies   have  opened new lines of research devoted, for example,
to the study of  supermembrane and Hamiltonian
\cite{M1,M2,M3,MR,FL,MOR,BR}.
\par In the context of string theory \cite{SW}, NC geometry
appears from the study of the quantum properties of D-branes in
the presence of non zero $B$-field. In particular, for large
values of $B$-field,
the spacetime worldvolume coordinates
do no longer commute and the usual product  of functions is
replaced by  the star product of Moyal bracket. In the context of
M-theory, NC structure is also present \cite{FL,MOR}.  For
instance, it can emerge from the L.C.G. Hamiltonian of the
supermembrane with fixed central charge. The central charge can be
induced by an irreducible winding around a flat even torus
\cite{MR,MOR,BR}. Using minimal immersions associated to a
symplectic matrix of central charges as backgrounds,   a non
commutative symplectic supersymmetric Yang-Mills  theory, coupled
to the scalar fields
transverse to the supermembrane,   has been  obtained. The
main physical interest of this NC formulation of the supermembrane
relies on the discreteness of its quantum
spectrum \cite{M2} and mainly \cite{M3} in clear contrast to the commutative
case \cite{LN,PP}.   NC geometry has been also used to study tachyon condensation
\cite{GMS}.  However, most of the NC spaces considered in all these studies involve mainly NC ${\bf R}_{\theta }^{d}$ \cite{GMS} and
NC tori T$_{\theta }^{d}$ \cite{SS}. \par Recently efforts
have been devoted  to go beyond these particular geometries  by considering  NC Calabi-Yau manifolds
used in string theory  compactifications.
A  great interest has
been given to build NC Calabi-Yau (NCCY)  threefolds using the  so-called
algebraic geometry method  introduced first by Berenstein and Leigh in  \cite{BL}.
The NC aspect  of this manifold is quite important for the stringy resolution of
  Calabi-Yau singularities and  the interpretation of string states.  In particular,   in string theory on Calabi-Yau manifolds,   NC deformation
is associated with open string states while the commutative  geometry  is in one to one
correspondence  with closed string states. This  formulation has been extended
  to higher dimensional manifolds  being  understood as homogeneous
hypersurfaces  in ${\bf CP^{n+1}}$ \cite{BS1}, or more generally
  to hypersurfaces in toric varieties \cite{BS2,BeS}.
\par  The  aim  of this paper  is to extend those
results to the case of  manifolds with exceptional holonomy
groups used in string theory compactifications. In particular,  our main objective is to develop an explicit analysis for
a NC  $G_2$ structure.   Using  an algebraic
orbifold method, we present NC  aspects of
manifolds with  $G_2$  holonomy group. We  develop and solve the
non commutativity  parameter  constraint  equations
defining    such structure algebras.
We   find   that there are eight possible solutions  for this extended structure,
  one of which corresponds to the commutative case.  Using  combinatorial arguments   we give
matrix representations solving such algebras. A matrix model of M-theory on  $G_{2}$ manifolds is found. Its generalization to NC $G_{2}$ is discussed.
\par  The outline
of the paper is as follows.  In section 2, we present
NC  aspects of   $G_2$ structure manifolds  using an algebraic
orbifold method.  In section 3,  we  develop and solve the
parameter constraint
equations
defining       such a structure. Then, we show that there are eight possible
solutions  for this extended  geometry. In section 4,    we give
matrix representations solving such algebras. An application to  matrix model
  of M-theory is discussed in section 5. Then, we give our
  conclusion  in section 6.
  We finish this work
by  an appendix.
\section{    NC $G_2$ Structure }\label{ade}
In this  section  we  want   to   present   a  non commutative  geometry  with  the  $G_2$  holonomy group.
This may extend  results  of the NC Calabi-Yau
geometries. First,  the $G_2$ structure
appears in  a seven real dimensional
manifold, with holonomy group $G_2$,  and  plays a  crucial
role in the M-theory  compactification. In particular, it was shown
that in order to get  four dimensional models  with only four
supercharges
   from M-theory,  it is necessary to consider a compactification
    on   such a
structure manifold.    Like  in 
Calabi-Yau   case, there are several realizations and   many non-trivial  $N=1$ models
in four dimensions   could be  derived, from  M-theory,
  once  a  geometric realization has been considered.  Before going ahead let us recall
what is the,  commutative,   $G_2$  structure.  Indeed,  consider a $ \bf  R^7$
parametrized  by $ (y_{1}, y_{2},...., y_{7})$. On this space,  one
defines   the metric
\be
g=dy_{1}^{2}+....+dy_{7}^{2}. \ee Reducing $SO(7)$  to  $G_2$,
one can define  also a special   real
  three-form  as  follows
\be
\varphi=dy_{123}+dy_{145}+dy_{167}+dy_{246}-dy_{257}-dy_{356}-dy_{347}\ee
where $dy_{ijk}$ denotes $dy_idy_{j}dy_{k}$. This expression of
$\varphi$   comes   from  the fact that $G_2$  acts as an
automorphism group  on the
octonion algebra  structure  given by
\be
t_it_j=-\delta_{ij}+f_{ij}^kt_k,  \ee  which   yields  the  correspondence
\be
f_{ij}^k  \to dy_{ijk}. \ee
The couple  $(g,\varphi)$  defines the so-called    $G_2$ structure. \\
In what follows,  we  want to deform the above structure  by
introducing  NC geometry.  This
deformation
may extend   results of NC Calabi-Yau  geometries studied in \cite{BL,BS1,BS2, BeS, KL,BMR}.
It  could be  also used to resolve   the  $G_2$    manifold  singularities by  non commutative  algebraic method.
\\
Simply speaking, the  $G_2$  structure   could be  deformed   by  imposing
the constraint
\be
y_iy_j\neq y_jy_i.
\ee
Basically, there are  several  ways  to  approach  such
a deformed geometry.  For instance, one may use  the  string theory approach
developed by Seiberg and Witten in \cite{ SW}. Another way, which we are interested in this present
work,  is to use an  algebraic geometry  method   based on solving the
non commutativity  in terms of discrete isometries of  orbifolds \cite {BL, BS1}.

\subsection{  Constraint equations of NC $G_2$  structure}

To get the constraint equations  defining  NC $G_2$  structure,
we proceed in steps as follows.  First,  we    consider  a
discrete symmetry  $\Gamma$, which will be specified  later on,
  acting   as follows
\bea \Gamma:\quad y_i\to \alpha_i y_i,\quad \alpha_i\in \Gamma.
\eea The resulting space is
constructed by identifying the points which are in
  the same orbit under the action of the group,
i.e., $y_i\to \alpha_i y_i$.  It
is smooth everywhere, except at the fixed points, which are
invariant under non trivial group elements of $\Gamma$.
  The invariance of  the  $G_2$ structure under $\Gamma$
    can define
a non compact seven dimensional
manifold with holonomy group $G_2$. The
compactification  of this geometry  leads to   models  studied by
  Joyce in \cite{J} \footnote{Other realizations of $G_{2}$ manifolds have been
developed mainly in the context of M-theory compactifications.}. Then,   we      see the orbifold space as a NC algebra.  We seek to  deform the
algebra of functions  on the
orbifold of $ \bf R^7$ to a NC
algebra ${\cal{A}}_{nc}$. In this way, the center of this
algebra is generated by the quantities invariant under the
orbifold symmetry.   In particular,  we  imitate the  BL orbifold
method given in \cite{BL}  to
  build a  NC extension of  $({{\bf R^{7}} /  \Gamma})_{nc}$. This extension  is obtained, as usual, by
extending the commutative algebra ${\cal{A}}_{c}$ of functions on
$\bf R^7$ to a
NC one ${\cal{A}}_{nc}\sim {\left({{{\bf R}}^7 / \Gamma}\right)}_{nc}$.
  The NC  version of
the orbifold ${{\bf R^{7}}}/{\Gamma }$ is obtained by
substituting the usual commuting $y_{i }$ by the matrix
operators $Y_{i }$   satisfying  the
following NC algebra structure \footnote{This algebra can be
viewed as the Yang-Baxter equations.}
\begin{eqnarray}
\label{NC} Y_{i }Y_{j } &=&\Theta_{ij } Y_{j }Y_{i},
\end{eqnarray}
where $\Theta$  is a matrix  with further  properties arranged
in such a way as to preserve the  $G_2$ structure.  In this way,
$\Theta_{ij}$   should   satisfy   some constraint equations
defining the explicit  NC  $G_2$ structure.  Here we want  to derive
  such parameter constraint relations.
To do so,  let us start by writing down  the trivial ones. Indeed, eq.(\ref{NC}) requires  that \be\label{ce1}
\Theta_{ij}\Theta_{ji}=\Theta_{ii}=1. \ee
However, non trivial relations   come from the
structure defining the commutative geometry. The   crucial property in our method   is that
  the entries of the  $\Theta$  matrix
must  belong to $\Gamma$, i.e
\be\label{ce2} \Theta_{ij} \in \Gamma. \ee
The  invariance of
the $G_2$ structure under  $\Gamma$  requires   that
\bea\label{ce3} \Theta_{ij}\Theta_{ik}&=&1\qquad 
\mbox{for}\quad i\neq j\neq k,
\quad i,j,k=1, \ldots,7 \\
\Theta_{ij}\Theta_{ik}\Theta_{i\ell}&=&1\qquad \mbox{for}\quad i\neq j\neq k\neq
\ell, \quad i,j,k,l=1, \ldots, 7. \eea
Let us remind that $\Theta_{ij}$ are coefficients of the matrix $\Theta$, so no summation on the indices is  considered.

\subsection{   Solving  the   constraint equations }
Before studying the corresponding  matrix representation, we
solve  first the above  parameter constraint equations.
This is needed to   define   explicitly
the NC  $G_2$ structure. It turns out that,  an explicit solution  can be
    obtained
once we know the elements of the center
  ${\cal Z}({\cal{A}}_{nc})$. The latter, which
  yields the commutative algebra  generated by
quantities invariant under the action of $\Gamma$,  is just the commutative $G_2$ geometry.
  It   may be a  singular  manifold
  while the geometry  corresponding to  the NC algebra will
   be a deformed one.
In other words, the commutative singularity can be deformed
in a NC  version of  orbifolds and can have a physical interpretation in M-theory
compactifications.\par
Let us  now   specify the   discrete  group
   $\Gamma$.  In order  not to loose  contact with the commutative case, consider $\Gamma$ as $Z_2\times Z_2\times Z_2$ \footnote{This could be extended to any discrete subgroup of  $G_2$ Lie group preserving
the $G_2$  structure.}.This symmetry
  has been studied in \cite{J}  to construct compact $G_2$  manifolds.
Since in this case,   $y_i^{2}$ is invariant under ${\Gamma }$,
  then the corresponding  operator  $Y_i^{2}$
should be in the center of the ${\cal Z}({\cal{A}}_{nc})$. This implies
  that \be \label{ce4} \Theta_{ij} \Theta_{ij}=1,\ee
which is consistent with the invariance of  the metric.  This  equation  is a strong constraint which will have a
  serious consequence on  solving   NC
$G_2$ structure.  (\ref{ce4}) can be solved by taking  $\Theta_{ij}$ as
\be\label{so1}
\Theta_{ij} = (-1)^{\epsilon_{ij}}.
\ee  Here  ${\epsilon_{ij}}$  is a matrix  such  that
${\epsilon_{ij}}+{\epsilon_{ji}}$
  is equal to zero   modulo $2$, which  is    required by (\ref{ce1}-\ref{ce3}).  A possible solution  is given by $\Theta_{ij} = (-1)$ where
   ${\epsilon_{ij}}=1$ modulo $2$, corresponding to  a flat space.  However, the   invariance of the
$G_2$ structure leads  to a solution
   where  some $\Theta_{ij}$  are equal to one. Using (\ref{NC}-\ref{so1}),  one   can  solve $\Theta_{ij}$ as follows
\begin{equation}\label{so2}
\Theta_{ij}=\left(
\begin{array}{ccccccc}
1 & a & a & b & b & c & c\\ a & 1 & a & d & e & d & e \\ a & a & 1
& f& g & g & f \\ b & d & f & 1& b & d & f  \\ b & e & g & b & 1 &
g & e \\ c & d & g & d & g & 1 & c \\ c & e &  f & f & e & c & 1
\end{array}
\right),
\end{equation}
where the  entries  of this matrix are integers such that \bea
a,b,c,d,e,f,g&=&\pm 1 \\ a&=&bdg  \\ a&=&bc=de=gf. \eea
Since the      representation  of the above algebra depends on this  matrix,
let us make  two  comments. First, 
the  algebra of  NC  $G_2$  structure contains
commutation and
anticommutation relations. Second, we find there are eight different solutions corresponding
   to the two different choices of one of the integers, namely  $ a=\pm 1$.  They are classified as follows
\begin{equation}
\label{tab}
\begin{tabular}{|l|l|l|l|l|l|l|}
\hline a&b&c&d&e&f& g\\ \hline 1&1&1&1&1&1&1 \\ \hline
1&-1&-1&-1&-1&1&1
\\ \hline
1&-1&-1&1&1&-1&-1
\\ \hline
1&1&1&-1&-1&-1&-1
\\ \hline
-1&-1&1&-1&1&1&-1
\\ \hline
-1&1&-1&1&-1&1&-1 \\ \hline -1&1&-1&-1&1&-1&1\\ \hline
-1&-1&1&1&-1&-1&1 \\ \hline
\end{tabular}
\label{tabo}
\end{equation}
From this classification, one can  learn   that we  have  eight different representations solving  the NC $G_2$  structure.
The trivial  one  corresponds  to all the parameters being equal
to $+1$,  which is equivalent
to have a complete set of commutative relations as a subset
  of possible solutions.

\section{  Matrix representation of NC   $G_{2}$ structure}\label{alg}

In this section,  we  construct  eight different representations $\{Y_{i}\}_{a,b,c..g}$
corresponding to the  above  NC  $G_2$ structure.   Our representation will be given
in terms of infinite dimensional matrices with the
following block structure
\begin{equation}
\label{matrix}
Y_{i}=\left(
\begin{array}{ccccc}
M_{i} & 0 & 0 & 0 & ... \\ 0 & M_{i} & 0 & 0 & ... \\ 0 & 0 &
M_{i} & 0 & ... \\ 0 & 0 & 0 & M_{i}& ...\\ \vdots &\vdots & \vdots &\vdots
&\ddots
\end{array}
\right), \quad i=1,\ldots,7.
\end{equation}
Here $M_{i}$,  which are  $2^{7}\times 2^{7} $ matrices,    satisfy the NC  $G_{2}$
structure given by (\ref{NC}) \footnote{Note that the size is related to
the number of generators. Details are  given in \cite{M3}.}. The    constraint
$\Theta_{ii}=1$  require that  $M_{i}$  should  be symmetric
matrices ( i.e $M_{mn}=M_{nm})$. \\
Our   way  to give  explicit representations   is  based on the
matrix  realization  of the grassmanian algebra of spinors $SO(7)$
in eleven dimensions found in \cite{M3} although there are some
differences in our case. The entries of the matrices $M_{i}$ are
$\{+1,-1,0\}$. The vanishing coefficients are the same as in the
symmetrized version of the matrix representation found in
\cite{M3}. However, the nonvanishing entries differ in their signs
respect to \cite{M3}. For each $M_{i}$, the signs are determined by the
NC  $G_{2}$ structure, through $\Theta_{ij}$ in the following niceway. Indeed, 
 let us
define a  vector $s_{i}$ that we shall call it  vector of signs as \bea
s_{i}=(+,\Theta_{1i})\otimes
\ldots\otimes(+,\Theta_{(i-1)i}),\quad i=1,\dots,7 \eea with this
product  defined as \bea (a_{1},\ldots,
a_{k})\otimes(b_{1},b_{2})\equiv(a_{1}b_{1},\ldots,a_{k}b_{1},a_{1}b_{2},\ldots,a_{k}b_{2}).
\eea
The $k$-th element of this vector has the following expression
\bea (s_{i})_{k}=\left\{\begin{array}{ll}
\displaystyle{\prod_{\ell=1}^{i-1}} \Theta_{\ell i}\quad &
2^{\ell-1}+1+2^{\ell}p\leq k\leq 2^{\ell}(p+1)\quad  p=0,\ldots,2^{i-(\ell+1)}-1\\
+          &   \textrm{otherwise}
\end{array}\right.
\eea
In terms of this $(s_{i})_{k}$ the matrices $M_{i}$ can be expressed as

\bea
(M_{i})_{mn}=\left\{\begin{array}{ll} (s_{i})_{k} &\quad
m=k+2^{i}v \quad\quad  n=m+2^{i-1}\\
            &\quad k=1,..,2^{i-1}\quad\quad  v=0,..,2^{7-i}-1\\
0           &\textrm{otherwise}
\end{array}\right.
\eea
Note that $(s_{i})_{k}$ for a given matrix $M_{i}$ is repeated over $v$.

It is easy to see that $M_{i}$ are hermitic traceless matrices of
$2^7\times 2^7$ size that satisfy the NC $G_{2}$ algebra. In order to
extend our results to  NC  $T^{7}/(Z_{2}^{3})_{\Theta}$ we can
construct eight infinite dimensional 
    representations $Y_{i}$ using  (\ref{matrix}) by considering that the periodic boundary
   condition is imposed at the $\infty$. Furthermore,  
our  representation  could solve a  general  algebra  satisfying
\be
U_{i}V_{j}=\Theta_{ij} V_{j}U_{i} \ee  where   $\Theta$ is a matrix
containing real roots of the identity.

\section {  Link  to  matrix model of M-theory }\label{alg}
In this section, we want to study  an application of NC $G_2$ structure in M-theory compactifications.
In particular, we discuss a possible link  to matrix model of M-theory. Before going ahead,
let us first review  such a  formalism  and then try to connect it  to our NC $G_2$ structure solutions.
Indeed, matrix model of M-theory is defined by maximally
supersymmetric $U(N)$ gauge quantum mechanics \cite{BFSS}.
In the
infinite momentum frame,  this dynamic  is described by the following   SYM lagrangian
\bea
S_{D0}=\frac{-1}{2gl_{s}}tr((\dot{X}_{i})^{2}+\frac{1}{2}[X_{i},X_{j}]^{2}+
\Psi^{t}(i\dot{\Psi}-\Gamma^{i}[X_{i}, \Psi]).\eea
Here $X_{i}$
are nine hermitian $N\times N$  matrices  representing the
transverse
coordinates to $N$ $D0$-branes in type IIA  superstring theory.   For no commuting transverse  coordinates,  this leads to   fuzzy  geometry in M-theory compactifications. \\
In what follows, we want to  interpret  seven of these  $X_{i}$ as
operators  satisfying   our   NC  $G_{2}$
structure,  while we take  $X_{8},X_{9}=0$.  In this way, the
vacuum equations of motion, for the static solutions,   read  as
\bea
\label{equation} \displaystyle\sum_{i}[X_{i},[X_{i},X_{j}]]=2
\displaystyle\sum_{i}(1- \Theta_{ij})X_{j}, \eea 
where  $X^{2N}_{i}={\bf I}_{N\times N}$.\\
A  simple  computation reveals that,  the
commutative solution  of  $G_{2}$ structure  solve immediately
(\ref{equation}). In this case the $X_{i}$
  matrices can be diagonalized and their $N$  eigenvalues represent the
  positions of the $N$ $D0$-branes. However the NC solutions  do  not satisfy (\ref{equation}). It is easy to see by using the result given in (\ref{tab}). Indeed,   for each case,  one   gets
\bea \displaystyle\sum_{i}[X_{i},[X_{i},X_{j}]]=2
\displaystyle\sum_{i}(1- \Theta_{ij})X_{j}=
\left\{\begin{array}{ll} \neq 0 &i\neq  j\\ 0 & i= j
\end{array}\right.
\eea
However, one can solve the equations of motion even for the NC  solutions by using
a physical modification. Before  doing that,   let us first  make a  comment.   We note
that a  singular characteristic of  NC $G_2$ structure  solutions corresponds to the case where  one
coordinate,  which  represents a direction
in the transverse space, commutes with the remaining ones \footnote{Our configuration for the non
commutative cases  $(R^{7}/Z_{2}^{3})_{\Theta}$ is acting as a $(R^{6}/Z_{2}^{3})_{\widetilde{\Theta}}\times R$ where
\begin{equation}
\Theta_{ij}=\left(
\begin{array}{ccccc}
1 & \dots & & ...&1 \\ \vdots & & & &   \\  &  &
\\  &  & & \widetilde{\Theta}_{i'j'} &\\ \vdots &  &  & & \\ 1 &  &  &
&
\end{array}
\right), \quad i,j=1,\ldots,7.\quad i',j'=2,\ldots,7
\end{equation}
For a solution with $[X_{1},X_{i'}]=0$ with $X_{1}$ coordinate of $R$ and the rest of the
solutions satisfying the algebra $X_{i'}X_{j'}=\widetilde{\Theta}_{i'j'}X_{j'}X_{i'}$. This
  is a general structure for all of our NC solutions.}. In this way, the seven different solutions represent
  the different possible choices of this direction. The  commutative coordinate represents a flat direction in the
   potential.   It turns out that  NC  solutions could  solve equations (\ref{equation})   if one
    introduces   higher energy corrections in the
D0-brane  action. This
generates   a constant quadratic contribution in  the action  which takes now  the following form  \bea\label{mass} S=
S_{D0}+ \mu_{j}A_{j}^{2}. \eea
Setting
$\mu_{j}=\frac{1}{4}\displaystyle\sum_{i}(\Theta_{ii}-\Theta_{ij})$,
(\ref{equation}) are now  satisfied. Indeed,  these  mass terms are directly induced by
anticommutative contributions  of  NC $G_{2}$ structure providing a massive
potential for six of the seven directions. For a given flat direction, let say  $X_{1}$,   we have
\bea \mu_{j}=\left\{\begin{array}{ll} 0   &j=1\\ 8   &j=2,\ldots,
7
\end{array}\right.
\eea To find a link  with the matrix model of M-theory for  NC
solutions,  we should find  explicit  terms  for  couplings leading  to
mass terms in the above  action,  being
related to NC $G_{2}$ structure in the
regularized models. Alternatively, mass terms
have  been present in matrix models which are a regularization
of theories containing also more contributions
in the action as cubic or quartic terms. This is a common fact when fluxes are
turned on in a theory as happens in Myers effect \cite{myers}, or in soft breaking terms \cite{angel}.
\par
On the other hand, the Hamiltonian of the supermembrane matrix model with non trivial central charge
on a two torus  has been  studied in \cite{M1}-\cite{M3}. It is a noncommutative symplectic
super Yang-Mills  coupled to transverse scalar fields of the
supermembrane. It contains quadratic, cubic and quartic contributions. However, if we restrict
  to the bosonic sector and fix the gauge field $A_{r}=0$, the regularized model is
  reduced to \bea
S_{D0}&=&\frac{-1}{g^{2}}tr(\frac{1}{4}[X^{m},X^{n}]^{2}+
(\widehat{\lambda}_{r}X^{m})^{2})\quad n, m=1,\ldots,7\\
\widehat{\lambda_{r}}^{A}_{B} &=&f_{r(B-r)}^{A} \quad
r=(1,0),(0,1).\eea This is the matrix model expansion of  \bea
\{\widehat{X}_{r},X^{m}\}=D_{r}X^{m}, \eea where $\widehat{X}_{r}$
are fixed backgrounds and minimal immersions of the supermembrane.
These backgrounds are responsible of the NC structure
of the supermembrane and they are associated to the non trivial
central charges \cite{M3,BR}. In spite of the formal analogy  between the structure of this
theory and our model,  there is an important difference. If we
interpret the $m$ transverse coordinates as the ones  satisfying our
NC $G_{2}$ structure and imposing
$\lambda_{r}=\lambda$,  the  mass term contribution has its
origin in a particular set of the structure constants $f_{AB}^{C}$
of $SU(N)$ although in our case the mass terms are directly fixed by the
NC  $G_{2}$ structure.
\par
In  type IIB superstring, a D-instanton matrix model
for a massive SYM, without extra terms, has  a fuzzy sphere and a
fuzzy torus as possible  solutions.
In this case,  the mass term is negative leading to some
instabilities,  although it
is free to be set to different values.
However, the origin of this extra term  is not well understood \cite{kimura}.
In our model even if we find a matrix model that allows  us to fix its mass term to  our
constant value, this mass coupling also remains unclear to us.

\section{ Conclusions}\label{alg}
In this  study, we have presented a  NC  $G_2$
structure  extending  results of  NC  Calabi-Yau manifolds.  In this way,  singularities of  $G_2$ manifolds  can be deformed by NC
algebras.
Using
a  algebraic  orbifold  method, we have given an explicit analysis for
building a NC $G_2$ structure.  In particular, we have  developed  and  solved the non commutativity
  parameter constraint  equations defining   such a  deformed geometry. Then,
we have   shown   that there are eight possible solutions  having similar
  features of Yang-Baxter equations.   Using a combinatorial argument,  we have found  eight
matrix representations  for such  solutions.\par Our results could
be extended to Spin(7)  holonomy manifold. The latter is  an eight
dimensional  manifold with Spin(7) holonomy  group, being a
subgroup of GL(8,$\bf R$)  which  preserves  a self dual 4-form
given by $ \varphi =dx_{1234}+dx_{1256}+dx_{1278}+dx_{1357}
-dx_{1368}-dx_{1458}-dx_{1467}-dx_{2358}-dx_{2367} -dx_{2457}
+dx_{2468}+dx_{3456}+dx_{3478}+dx_{5678}$.  The invariance of such
a form, under  a  discrete group of spin(7),  leads to non trivial constraint equations defining NC
$Spin(7)$  manifolds.
\par
A matrix model of M-theory has been found for the commutative solution
of the  $G_{2}$ structure. It is a particular case of the deformed one.
It satisfies trivially the
solution  to  the vacuum equation
of motion. However, this is not the situation for the NC   geometries. We have shown that higher energy
corrections  are needed  to satisfy  such equations. These   extra  quantities being
  mass terms give information about NC  structure. We argue that they appear as a
   consequence of the resolution of the singularity by introducing a NC  algebra.
The explicit coupling that leads to  these mass terms in the regularized matrix model,
remains unclear  to  us.
  In order to find  a complete  connection with matrix model for  NC  solutions a more
   extensive analysis would be required.\\ On the  other hand,
        a paper \cite{patil}
     dealing with
       topological transitions in fuzzy spaces  has been  appeared recently. It also involves
        mass terms  producing  a topological change. Its suggested origin
       is a Yukawa interaction term in the D0-brane  action. It has certain resemblance
        with our case  studied here, although  we do not know if there could be an underlying
        relation between both approaches.
        We leave these open questions
   for future work.

\section*{Appendix} In this appendix, we
want to show  that our  eight representations satisfy eq. (\ref{NC}). To do so,    we should prove the following
constraint equations
\bea M_{i}^{2}&=&{\bf I}\\
M_{i}M_{j}&=&\Theta_{ij}M_{j}M_{i}\quad i\ne j \eea  First, let
us   denote a nonzero coefficient  $(a_{i})_{mn}$, of the matrix
$M_{i}$,   by the position that occupy in rows and columns
   $(m,n)_{(s_{i})_{k}}$, where ${(s_{i})_{k}}$ its sign.
\\
$(i)$ $M_{i}^{2}={\bf I}$\\ It is easy to show this property. Indeed, from the construction of the
matrices, we  can see that there are no repeated coefficients and
there is  only one nonzero coefficient per row or column.
  Since  the  matrices are symmetric and traceless,  the product between one term
   and its adjoint is the only contribution per line to the diagonal. For a given matrix $M_{i}$,
    the signs depend  only on $k$ which is the same for both type of terms.
     The product of signs is then the square of each sign. Since they are
      real  square roots of the identity,  we have
\bea (M_{i})_{mn}^{2}=(m,n)_{(s_{i})_{k}}(n,m)_{(s_{i})_{k}}=
(m,m)_{+}={\bf I} \eea $(ii)$
$M_{i}M_{j}=\Theta_{ij}M_{j}M_{i}\quad i\ne j$.\\ In order to
prove this statement we need to check  the followings
\\
(1)  The nonzero coefficients of $M_{i}M_{j}$ are the same to the
ones of $M_{j}M_{i}$.\\ (2) The relative sign between each of the
coefficients of the two products is $s_{ij}=\Theta_{ij} s_{ji}$.
\\
In what follows we suppose $i<j$ without any lack of generality.
The matrix $M_{i}$ can be expressed  in terms of the nonzero
coefficients as $\{(a_{i})_{mn}, (a^{\dag}_{i})_{mn}\}\quad
\textrm{for}\quad n>m$. The product of coefficients  takes  then
the following  form \bea
(a_{i}a_{j})_{mo}=(a_{i})_{m(m+2^{i-1})}(a_{j})_{(m+2^{i-1})(m+2^{j-1}+2^{j-1}= o})
\eea
In an abbreviate notation, the product of the two matrices is
given by \bea (M_{i}M_{j})=\{a_{i}a_{j},a_{i}a^{ \dag}_{j},a^{
\dag}_{i}a_{j},a_{i}^{ \dag}a_{j}^{ \dag}\}.\eea
If this relation
is proved for $a_{i}a_{j}$ and $ a_{i}a^{\dag}_{j}$, then  their
adjoint terms will also satisfy it. Let  us  first deal with
$a_{i}a_{j}$. Indeed, the terms that contribute in the computations  are \bea
a_{i}a_{j}:(m,n)_{(s_{i})_{k}}(n,o)_{(s_{j})_{l}}=
(m,o)_{s_{ij}}\\
a_{j}a_{i}:(m,r)_{(s_{j})_{l}}(r,o)_{(s_{i})_{q}}=(m,o)_{s_{ji}}.
\eea (1) We are not going to care about the signs.
\\ Given $(a_{i})_{mn}$ if there exists a
$(a_{j})_{no}$, then one can find  $(a_{j})_{mr}$ and
$(a_{i})_{ro}$ such that
$(a_{i})_{mn}(a_{j})_{no}=(a_{j})_{mr}(a_{i})_{ro}$.\footnote{Only
for a given $(a_{i})_{mn}$ with $v=0$ and $i\le j$, it  is
guaranteed that $(a_{j})_{no}$ exists.}
\\
(a)  Moreover, we can find  $(a_{j})_{mr}$  once  a nonzero coefficient exists in the
line  $m$, as there is just one per line  or column. In this way, we have
$(a_{i})_{mn}$, $0<m\le 2^{i-1}$ and  $0<m<2^{j-1}$ as required
for any matrix labeled by $j$. Then,  the term exists and by
definition is given by  $r=m+2^{j-1}$. \\(b) On the other hand,  $(a_{i})_{ro}$ exists if there is a
coefficient in the line  $r$ of the matrix $a_{i}$ satisfying
$r=k+2^{i}v^{\prime}$ for some $v^{\prime}$.  This implies that
$m=k+2^{i}v_{u}$.  Using  (a),   we have
$r=k+2^{i}v_{x}\quad\textrm{for}\quad v_{x}=v_{u}+2^{j-(1+i)}$, so
the term also exists and by definition $o=r+2^{i-1}$. In fact,
given $(a_{i})_{mn}$, $(a_{i})_{ro}$ is a coefficient with the
same $k$  translated in $2^{j-(i+1)}$ units of $v$.  For instance,
the product can be represented as in \cite{M3}, by \bea
ij:m\stackrel{2^{i-1}}{\rightarrow}n\stackrel{2^{j-1}}{\rightarrow}o\\
ji:m\stackrel{2^{j-1}}{\rightarrow}r\stackrel{2^{i-1}}{\rightarrow}o
\eea
Since  (a) and  (b)  are  verified,  then we have  \bea
(a_{i})_{mn}(a_{j})_{no}=(a_{j})_{mr}(a_{i})_{ro}. \eea
To prove (2) note that the relation between the signs is $s_{ij}=\Theta_{ij}s_{ji}$,  being equivalent to verify that
\bea (s_{i})_{mn}(s_{j})_{no}=\Theta_{ij}(s_{j})_{mr}(s_{i})_{ro}.
\eea For a given matrix, the signs  depend  only on $k$. This means
that any value in $j$ for a given $k$ has the same sign. Then, we have
   \bea (s_{i})_{mn}=(s_{i})_{ro}. \eea
It remains  to describe  $(s_{j})_{no}$
in terms of $(s_{j})_{mr}$. By definition
$s_{k_{i}}=\prod_{\ell\prime}\Theta_{\ell\prime i}$, then we should find  \bea \prod_{\ell\prime}\Theta_{\ell\prime
i}=\Theta_{ij}\prod_{q\prime}\Theta_{q\prime j} \eea 
However we do
not need to know the explicit decomposition in terms of
$\Theta$. To obtain the relation between the two signs,  it
is enough to know their relative values of $k$. We will denote by
$k_{n}$ the value associated to the sign $(s_{j})_{no}$ and
respectively $k_{m}$ to $(s_{j})_{mr}$. Since $(a_{i})_{mn}$ is given, the relative  difference between $n$  and $m$ is known and
we have  \bea n=m+2^{i-1}\\ k_{n}^{j}=2^{i-1}k_{m}^{j}. \eea
Since $j>i$, and from the definition of sign product, one can
check that $\Theta_{ij}$ changes in $s_{j}$ in each $2^{i-1}$
alternating values of $k_{j}$. So,  we have the following   \bea
(s_{j})_{no}=\Theta_{ij}(s_{j})_{mr}. \eea  It follows that    \bea
s_{ij}=\Theta_{ij}s_{ji}. \eea The same results can be obtained
for $(a_{i}a_{j}^{\dag})$ after  making minor changes to the
argument. Using  (1) and (2), for any $i$ and $j$,  one gets  \bea
Y_{i}Y_{j}=\Theta_{ij}Y_{j}Y_{i}.\eea
\section*{Acknowledgments}
We thank   J. Bellor\'{\i}n, A. Font,  C. G\'omez, C. Herranz, K.
Lansteiner, E. Lopez, J. Rasmussen,  E. H. Saidi, A. Sebbar,  A. Uranga, for discussions,
collaborations and  support.   A.B.   is supported by the Ministerio de
Educaci\'{o}n, Cultura y Deportes (Spain) grant SB 2002-0036.
M.P.G.M. is supported by a postdoctoral grant of the
Consejer\'{\i}a de Educaci\'{o}n, Cultura y Deportes de la Comunidad Aut\'{o}noma de la Rioja, (Spain).

\end{document}